# Experimental observation of strong light-matter coupling in ZnO microcavities: influence of large excitonic absorption

F. Médard[1], J. Zuniga-Perez[2], P. Disseix[1], M. Mihailovic[1], J. Leymarie[1], A. Vasson[1], F. Semond[2], E. Frayssinet[2], J.C. Moreno[2], M. Leroux[2], S. Faure[3], T. Guillet[3].

[1] LASMEA, UMR 6602 UBP/CNRS, 24 Avenue des Landais, 63177 Aubière Cedex, France.

[2] CRHEA-CNRS, Rue Bernard Gregory, Sophia Antipolis, 06560 Valbonne, France.

[3] GES, UMR 5650 CNRS, 34095 Montpellier Cedex 5, France.

**Abstract:** We present experimental observation of the strong light-matter coupling regime in ZnO bulk microcavities grown on silicon. Angle resolved reflectivity measurements, corroborated by transfer-matrix simulations, show that Rabi splittings in the order of 70 meV are achieved even for low finesse cavities. The impact of the large excitonic absorption, which enables a ZnO bulk-like behavior to be observed even in the strong coupling regime, is illustrated both experimentally and theoretically by considering cavities with increasing thickness.

During the last decade planar microcavities (MCs) have been extensively studied due to their potential to change the nature of the light-matter interaction in the so-called strong coupling regime (SCR) [1]. The SCR is reached when the cavity photonic mode is resonant with an excitonic level of the active layer leading to the formation of cavity polaritons (CPs) [2]. Their dispersion displays two branches which become exciton-like or photon-like far from the resonance [3]. CPs are admixed, half-photon and half-exciton, quasi-particles whose bosonic and non-linear properties are attracting much attention. The SCR manifested by an anticrossing of the exciton and photon modes was observed for the first time by *Weisbuch et al.* in a GaAs-based microcavity [4].

From the fundamental point of view, MCs constitute an appropriate system to manipulate polaritons and, thereby, to get further insight into the physics of the light-matter interaction. From the technological point of view, MCs operating in the SCR offer also the possibility to achieve bosonic effects and are thus promising candidates to investigate novel devices such as polariton light emitters [5,6] or optical amplifiers [7]. Moreover, electrically driven polariton emission in GaAs microcavities has been reported recently up to 235 K [8] and opens the way toward new sample designs suitable for the realization of electrically pumped polariton emitters. However, in order to raise the working temperature of these polariton-based devices up to room temperature, alternative semiconductors, presenting large exciton binding energies and oscillator strengths, should be employed. In this sense, III-nitrides based microcavities have thus been investigated, and large Rabi splittings, varying between 30 and 50 meV [9-13], have been reported. Another wide-bandgap material suitable for the realisation of polariton devices is ZnO because of its record excitonic oscillator strength [14,15] which should lead to larger Rabi splittings [14,16,17].

In this letter, we present experimental evidence of the SCR in bulk ZnO MCs and highlight the peculiarities of their reflectivity spectra due to the strong excitonic absorption. We study two hybrid ZnO MCs with bottom 7.5 period $Al_{0.2}Ga_{0.8}N$/AlN distributed Bragg reflectors (DBR) grown directly on Si(111) substrates and 10 nm aluminum top mirrors. In order to investigate the influence of the excitonic and band-to-band absorption in the active layer, the ZnO thickness has been varied from λ/4 to 3λ/4. The reflectivity of the bottom and top mirrors (0.84 and 0.54 respectively) leads to a quality factor (Q) which does not exceed 50 in each case. The nitride DBRs and the ZnO cavities where both grown by molecular beam epitaxy in two separate reactors. These samples were studied by reflectivity as a function of incidence angle (from 5° to 75°) at 5 K, 77 K and room temperature.

Angle resolved reflectivity spectra of the λ/4 cavity at 5 K are displayed on figure 1 for transverse magnetic (TM) polarization. Similar results have been obtained for the transverse electric (TE) polarization. At the lowest incident angle (5°) the lower polariton branch (LPB), which is mainly photonic, is observed at 3.258 eV and as the incidence angle increases it moves towards higher energies and approaches asymptotically the free A and B exciton average energy which is found to be 3.375 eV. Both excitonic features cannot be resolved due to a significant inhomogeneous broadening. Together with the increase of the LPB energy a reflectivity dip, located on the high-energy side of A and B excitons, moves toward higher energies and is attributed to the upper polariton branch (UPB). Between 35° and 55° it is further split in two components as figure 1 shows. The angle dependence of the measured reflectivity dips is given at the bottom of figure 1. The splitting of the UPB reflectivity feature at 3.44 eV can be explained by the strong increase of the ZnO imaginarypart of the dielectric function at this energy which is attributed to A and B exciton excited states, to the onset of continuum absorption and to exciton-phonon complexes [16,18]. This point will be further discussed in the part dedicated to the simulations. The Rabi splitting measured from the analysis of the reflectivity spectra is found to be 75±10 meV.

The reflectivity of this λ/4 cavity has also been studied as a function of incident angle at room temperature in order to assess the preservation of the SCR up to 300 K (not shown). Strong exciton-photon coupling is seen to persist and a Rabi splitting of 50±10 meV at 300K has been estimated, the resonance being attained at an incidence angle of 38°.

In contrast with the previous λ/4 cavity, the reflectivity of the 3λ/4 cavity exhibits a different behavior as a function of incidence angle. The spectra are reported on figure 2 together with the angular evolution of the main dip energies. A salient observation in these spectra is that the A and B excitonic features respectively located at 3.371 and 3.379 eV (vertical dotted lines on figure 2) are well resolved but are absolutely insensitive to the incident angle. As the incidence angle increases, the energy of the lower polariton branch (LPB) increases up to the A exciton energy but its amplitude is progressively reduced although the two excitonic features remain visible. From this observation, it is clear that the excitonic absorption is so strong that the cavity is no more efficient when the photon energy is equal to that of A or B excitons. In other words, the sample can be seen in the vicinity of the excitonic energy as a semi-infinite medium. Indeed the product $\alpha.d$ where $\alpha$ is the excitonic absorption coefficient and $d$ the ZnO thickness is ~ 4.5, much greater than unity. Nevertheless, on the high energy side of A and B, a weak dip appears for an incidence angle close to 15°, shifting to higher energy with increasing angle. This feature is identified as the UPB resulting from the exciton-photon coupling when the excitonic absorption is not intense enough to completely extinguish the photonic mode. Due to the inhomogeneous broadening, the exciton can exist away from the energy corresponding to the maximum of absorption.

Even if the optical mode is rather broad, a light-matter coupling persists with a Rabi spitting estimated to 70 ± 10 meV. In TM configuration, the UPB energy saturates at ~ 3.43 eV for large incidence angles, presumably due to interaction with the numerous levels existing in the 3.42-3.45 eV energy range [18], which are also responsible for the anticrossing observed near 3.44 eV in the λ/4 cavity (figure 1). In TM polarization, the oscillator strength of C exciton becomes stronger as the incidence angle increases, since it is strongly allowed when the electric field is parallel to the c axis [19]. Accordingly, we observe the C exciton reflectivity feature for incidence angles greater than 30° at 3.426 eV in TM configuration (figure 2). On the other hand it is not seen in TE configuration, as shown by the 35° and 45° TE reflectivity spectra added in figure 2. The detailed interactions of polaritons with the numerous levels of varying anisotropy present in this range are beyond the scope of this letter.

From the examination of these two samples, it is clear that, owing to the high excitonic absorption of ZnO, the design of the cavity is crucial to clearly observe a strong light-mater coupling. A specific care has to be taken on the thickness of ZnO bulk microcavities in order to detect the UPB in reflectivity or luminescence.

To analyze our experimental results and get further insight into the effects of ZnO huge excitonic absorption, simulations of our experimental results have been carried out within the transfer-matrix formalism. In order to obtain a reliable optical index for ZnO in the near bandgap region, the respective contributions of excitonic and band-to-band transitions to the dielectric function are evaluated separately. Outside the excitonic energy range the complex refractive index is obtained from our previous ellipsometry and reflectivity measurements performed on bulk ZnO layers [16]. The contribution of the excitonic transitions to the dielectric constant is modelled through Lorentz oscillators [20]. Only the three first excitonic states (n = 1, 2 and 3) of A and B excitons have been included in the calculation. C exciton is not taken into account in the simulation because of the difficulty to adjust its oscillator strength which varies in TM configuration. The knowledge of ZnO dielectric function allows simulating the evolution of reflectivity spectra for both samples as a function of incidence angle. Figure 3 shows the simulated reflectivity spectra of the λ/4 cavity. It is worth noting the same evolution of the lower polariton branch as in the experimental spectra: a nearly pure cavity-mode at 5° with an energy equal to 3.257 eV and, at higher angles, a marked dip stopped by excitons. The upper polariton branch starts with a quasi excitonic behavior and progressively moves towards higher energies. The excitonic energies and corresponding oscillator strengths of A and B excitons included in the simulation are reported in table 1. These oscillator strength values are close to those reported in previous works [14,15,19]. The simulation confirms also the large excitonic broadening since a broadening parameter of 30 meV is used in order to account for the experimental spectra. This leads to a large polaritonic mode reinforced by the low quality factor of the cavity. In addition, in the experimental spectra, a second coupling occurs at higher energy (3.43-3.45 eV) between the cavity mode and interacting states (A and B excited states, C exciton in TM polarization and exciton-phonon complexes) existing in ZnO around this energy [18]. It is worth noting that this anticrossing happens at approximately 75 meV above the free exciton energy, which is close to the longitudinal optical phonon energy in ZnO, but higher than those of the A and B first excited states and of the C. This behavior is also related to the increase of the complex refractive index of ZnO near the band gap (3.43eV).

Figure 4 shows the calculated reflectivity spectra of the $3\lambda/4$ cavity. The lower polariton branch (LPB) is moving from 3.307 eV (5°) to 3.365 eV (75°) just below the exciton energy. Due to the ZnO thickness ($3\lambda/4$), absorption has a significant influence on the upper polariton branch (UPB). Above the bandgap, this UPB becomes less pronounced. It is interesting to artificially remove the band-to-band absorption in the model: the spectra are not modified below 3.38 eV, but the UPB, indicated as solid squares in figure 4, is recovered as a pronounced reflectivity dip, which could lead to a Rabi splitting of 80 meV. We can therefore conclude that in this $3\lambda/4$ cavity, the LPB is a well-defined polariton state, but that the UPB is damped. Between LPB and UPB dips and in accordance with the experimental data, the reflectivity spectra exhibit two peaks corresponding to A and B excitons with angle independent energies. The high oscillator strengths used in the calculation (see table 1) corroborate the importance of the excitonic absorption near 3.38 eV. The broadening is found to be weaker than in the $\lambda/4$ cavity with a value of 12.5 meV for each exciton. In order to underline the influence of the excitonic absorption on the optical properties of ZnO microcavities, simulations of reflectivity spectra have been performed for three ZnO MCs with different thicknesses ($\lambda/4$, $3\lambda/4$ and $5\lambda/4$). Calculations have been carried out for a photonic mode energy equal to the mean energy of A and B excitons; they are displayed on figure 5. For the thinnest cavity, three polaritonic branches (LPB, MPB and UPB) are clearly in evidence and demonstrate the strong coupling regime. On the contrary, pure excitonic reflectivity features ($X_A$ and $X_B$) persist in the calculated spectra related to the $3\lambda/4$ and $5\lambda/4$ cavities. For A and B excitonic energies where the absorption is maximum, the cavity is no more efficient since the light cannot reach the bottom of the cavity and consequently cannot be reflected. It can be concluded that the thickness of the ZnO active layer has to be finely adjusted in order to clearly observe the strong light-matter coupling for both the LPB and UPB.

In summary, evidence for exciton-photon strong coupling has been reported in ($\lambda/4$ and $3\lambda/4$) ZnO bulk microcavities grown by molecular beam epitaxy on a silicon substrate. The strong coupling, with Rabi energies of the order of 70 meV, is achieved in both cavities even if their quality factor is smaller than 50. These results are promising for the observation of polaritonic non linear effects in ZnO microcavities but have also shown the impact of the strong excitonic absorption in ZnO. Indeed, for the $3\lambda/4$ microcavity, and when the energy is close to the exciton one, the cavity effect is lost due to the strong excitonic absorption. Furthermore, in this case the UPB is damped. In order to enhance the figure of merit of the observed strong coupling, it is now important to improve the finesse of the optical mode by the deposition of a dielectric DBR in place of the aluminum mirror and to find the best trade-off between the huge excitonic oscillator strengths in ZnO, the thickness of the active layer and the Q factor of the cavity leading to higher splitting-to-linewidth ratios.

This work is supported by the ANR ZOOM project, contract N° ANR-06-BLAN-0135.

## References

1. A. Kavokin, J. Baumberg, G. Malpuech, and F. Laussy, in *Microcavities* (Oxford University Press Inc., New York, United States, 2007).

2. A. Tredicucci et al., Phys. Rev. Lett. **75,** 3906 (1995).

3. R. Houdré et al., Phys. Rev. Lett. **73,** 2043 (1994).

4. C. Weisbuch, M. Nishioka, A. Ishikawa, and Y. Arakawa, Phys. Rev. Lett. **69,** 3314 (1992).

5. G. Malpuech et al., Appl. Phys. Lett. **81,** 412 (2002).

6. G. Malpuech, A. Kavokin, A. Di Carlo, and J.J. Baumberg, Phys. Rev. B **65,** 153310 (2002).

7. P. G. Savvidis et al., Phys. Rev. Lett. **84,** 1547 (2000).

8. S. I. Tsintzos, N.T. Pelekanos, G. Konstantinidis, Z. Hatzopoulos, and P.G. Savvidis, Nature **435,** 372 (2008).

9. N. Antoine-Vincent et al., Phys. Rev. B **68,** 153313 (2003).

10. I. R. Sellers et al., Phys. Rev. B **73,** 033304 (2006).

11. G. Christmann, R. Butté, E. Feltin, J.F. Carlin, and N. Grandjean, Phys. Rev. B **73,** 153305 (2006).

12. R. Butté et al., Phys. Rev. B **73,** 033315 (2006).

13. I. R. Sellers et al., Phys. Rev. B **74,** 193308 (2006).

14. M. Zamfirescu, A. Kavokin, B. Gil, G. Malpuech, and M. Kaliteevski, Phys. Rev. B **65,** 161205(R) (2002).

15. S. F. Chichibu, T. Sota, G. Cantwell, D.B. Eason, and C.W. Litton, J. Appl. Phys. **93,** 756 (2003).

16. M. Mihailovic et al., Opt. Mater.doi:10.1016/j.optmat.2007.10.023 (2008).

17. R. Shimada, J. Xie, V. Avrutin, Ü. Özgür, and H. Morkoç, Appl. Phys. Lett. **92,** 011127 (2008).

18. W. Y. Liang and A.D. Yoffe, Phys. Rev. Lett. **20,** 59 (1968).

19. A. Teke et al., Phys. Rev. B **70,** 195207 (2004).

20. L. C. Andreani, in *Confined Electrons and Photons* Edited by E. Bunstein and C. Weisbuch, ( New York, 1995), p.57.

**Table 1: Parameters for A and B excitons used for the simulations.**

|  |  | λ/4 cavity | 3λ/4 cavity |
|---|---|---|---|
| Energy (eV) | $E_X^A$ | 3.368 | 3.370 |
|  | $E_X^B$ | 3.376 | 3.378 |
| Polarizability | $4\pi\alpha_A$ | $1.1\ 10^{-2}$ | $1.0\ 10^{-2}$ |
|  | $4\pi\alpha_B$ | $1.2\ 10^{-2}$ | $1.7\ 10^{-2}$ |
| Broadening (meV) | $\Gamma_{A,B}$ | 30 | 12.5 |

**Figure captions**

**Figure 1:** *Upper part:* low temperature (T=5K) reflectivity spectra of a λ/4 ZnO bulk microcavity recorded for various incident angle from 5° to 75° under TM polarization. *Lower part:* the lower polariton branch (LPB) and upper polariton branch (UPB) energies are reported as a function of the incident angle (vertical axis).

**Figure 2**: (Color online) *Upper part:* low temperature (T=5K) reflectivity spectra of a 3λ/4 ZnO bulk microcavity recorded for various incident angle from 5° to 75° under TM polarization. In order to underline the influence of C exciton on the UPB, TE polarization spectra have been reported for 35° and 45°. *Lower part:* the lower polariton branch (LPB) and upper polariton branch (UPB) energies are reported as a function of the incident angle (vertical axis) for TM (full circles) and TE (open circles) configurations.

**Figure 3**: Simulated reflectivity spectra of the λ/4 cavity for various incident angles.

**Figure 4**: Simulated reflectivity spectra related to the 3λ/4 cavity. Solid squares correspond to the position of the UPB evaluated by removing the band-to-band absorption in the model.

**Figure 5**: (Color online) Calculated reflectivity spectra of ZnO bulk microcavities in the excitonic energy range for three ZnO thicknesses (λ/4, 3λ/4 and 5λ/4).

**Figure 1**

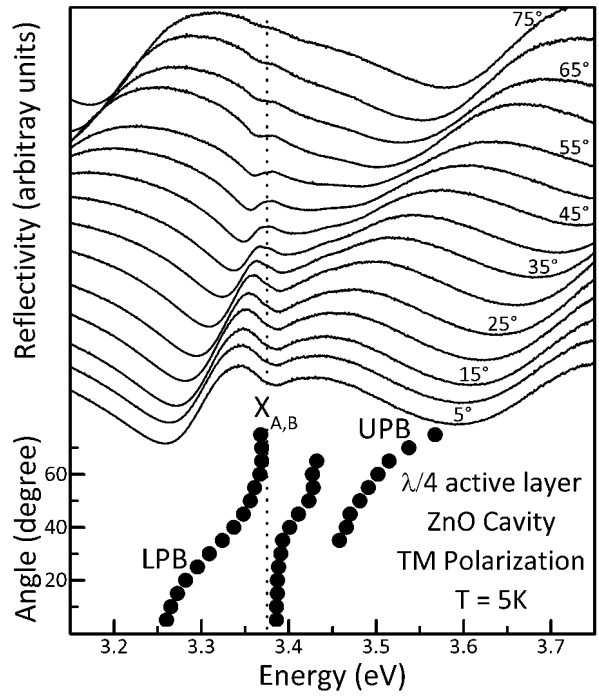

**Figure 2**

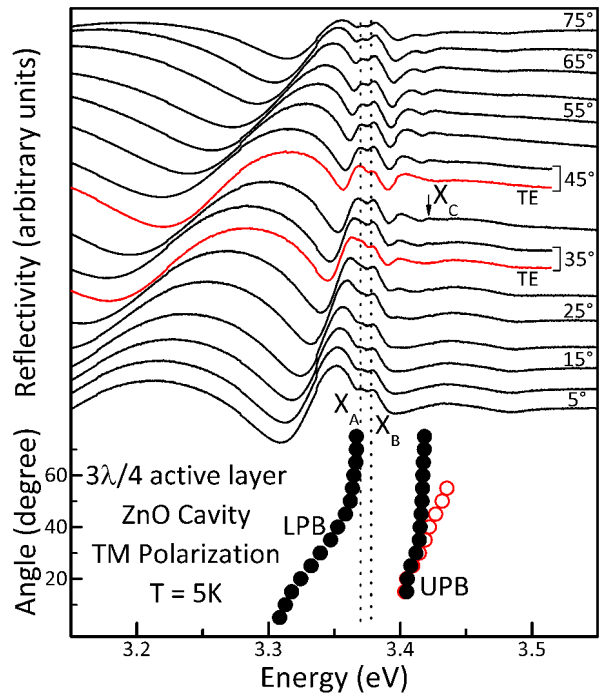

Figure 3

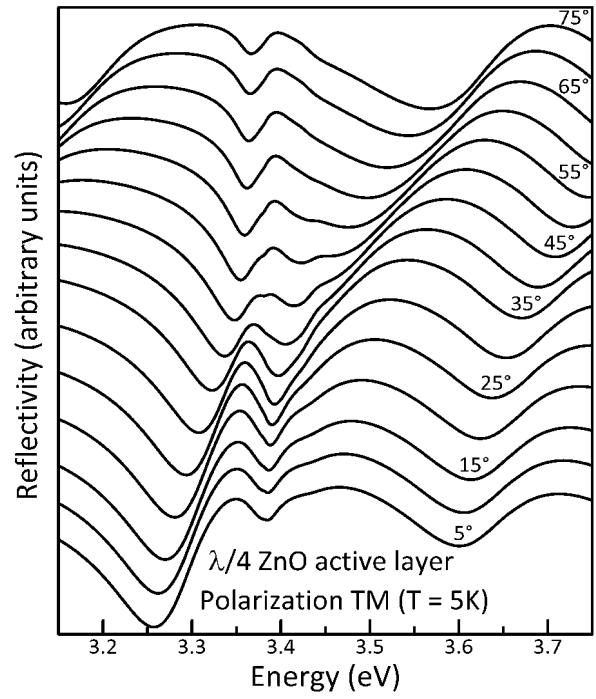

Figure 4

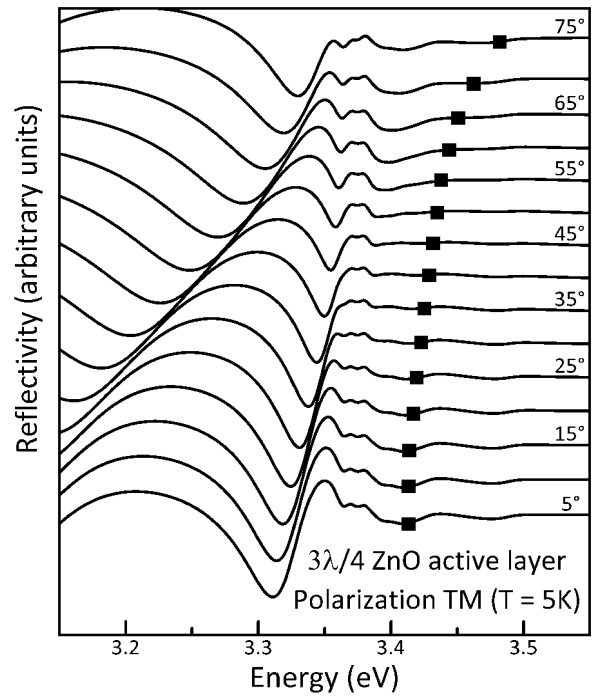

Figure 5Figure 5

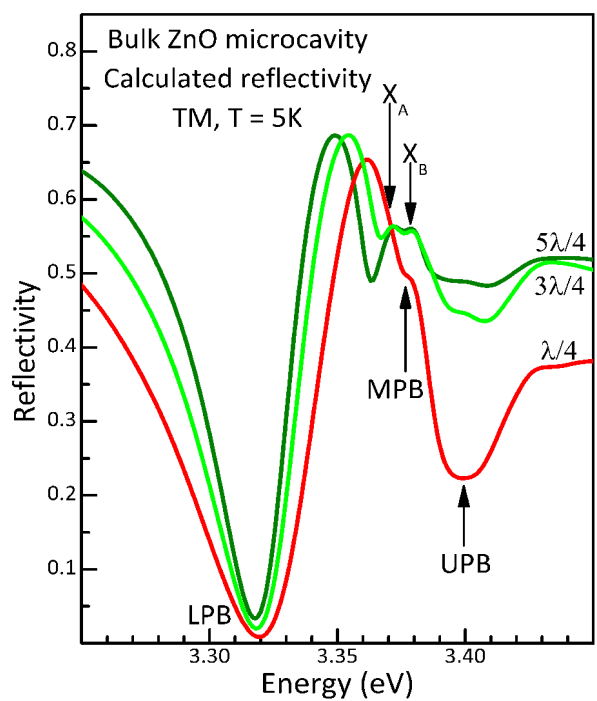